\documentclass[aps,twocolumn,amsmath,amssymb]{revtex4}
\usepackage{graphicx}
\begin{document}
\title{Practical fast gate rate InGaAs/InP single-photon avalanche photodiodes}
\author{Jun~Zhang}
\email{Jun.Zhang@unige.ch}
\author{Rob~Thew}
\author{Claudio Barreiro}
\author{Hugo~Zbinden}
\affiliation{Group of Applied Physics, University of Geneva, 1211 Geneva 4, Switzerland}
\date{\today}

\begin{abstract}
We present a practical and easy-to-implement method for high-speed near infrared
single-photon detection based on InGaAs/InP single-photon
avalanche photodiodes (SPADs), combining aspects of both sine
gating and self-differencing techniques. At a gating frequency of 921\,MHz
and temperature of -30\,$^{\circ}$C
we achieve: a detection efficiency of 9.3\,\%,
a dark count probability of 2.8$\times10^{-6}$\,ns$^{-1}$, while the
afterpulse probability is 1.6$\times10^{-4}$\,ns$^{-1}$, with a 10\,ns
``count-off time'' setting. In principle, the maximum count rate of the SPAD can
approach 100\,MHz, which can significantly improve the performance for
diverse applications.

\end{abstract}
\maketitle

InGaAs/InP SPADs provide one of the most important approaches for near infrared
single-photon detection, especially for practical applications
such as quantum key distribution (QKD) \cite{GRTZ02}.
Gated-mode InGaAs/InP SPADs have been well studied and
recently this has been extended to free-running mode \cite{TSGZR07,ZTGGZ09,WIB09}.
However, due to the afterpulsing effect and the need for tens of $\mu$s
deadtime settings, the gating frequency and
the maximum count rate of InGaAs/InP SPADs are both severely limited. On
the other hand, the requirements for long-distance and high bit rate
QKD systems motivate the development of high-speed near
infrared single-photon detectors, of which superconducting
single-photon detector (SSPD) \cite{SSPD} and up-conversion detector
(UCD) \cite{UCD} are two common candidates. Unfortunately, the
cryogenic requirements of SSPDs and the spurious nonlinear noise of
UCDs make them impractical for QKD systems. So far, two
types of high-speed InGaAs/InP SPADs have been demonstrated using the
techniques of sine gating (SG) \cite{NSI06,NAI09} and
self-differencing (SD) \cite{YKSS07,DDYSBS09}, respectively. These new
approaches have been demonstrated in QKD \cite{NFIHT07,QKD1,QKD2}
and random number generators \cite{RNG},
and have also shown photon-number resolving \cite{PNR}.

The afterpulsing effect is one of the major bottlenecks limiting the
performance of InGaAs/InP SPADs. The origin of afterpulsing
is due to the trapping of charge carriers by defects in the SPAD's
multiplication layer. Subsequent gates release some of these charges that
then create avalanches. The afterpulsing effect is not only attributed to the
defect concentration in the multiplication layer, which depends on
the impurity and device structure. It is also proportional to the
total number of carriers in an avalanche, which depends on the excess bias
of the SPAD and the avalanche duration time \cite{ZTGGZ09}.
In conventional gating, using the relatively long
gating time, the avalanche amplitude is large enough to be easily discriminated.
However, long deadtime settings are necessary to suppress the afterpulsing.

\begin{figure}[t]
\centering
\includegraphics[scale=0.76]{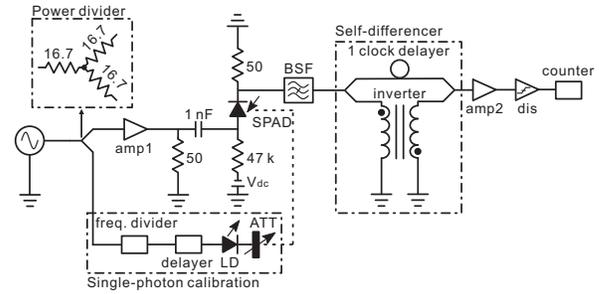}\\
\caption{The experimental setup.}
\label{sine}
\end{figure}

Conversely, in the case of rapid gating,
with gating frequencies ($f_{g}$) of around 1\,GHz, the ultra
short gating time ensures that the avalanches are far from saturation
and therefore the avalanche currents are quite weak. As such, the
afterpulsing effect can be significantly suppressed, although
with the disadvantage that the avalanche amplitudes are very
faint, normally a few mV, and thus difficult to discriminate.
If no avalanche occurs during a gate, the SPAD still outputs a
capacitive response, the background signal. In the case of
an avalanche the faint avalanche signal is superposed with this
background signal. Hence, the central task of rapid gating is
minimizing the background level to obtain enough single-to-noise ratio
to discriminate the small avalanche.

The SG method \cite{NSI06,NAI09} uses sine waves to gate a SPAD and
band-stop filters (BSFs) to filter out the background frequency
response, while SD \cite{YKSS07,DDYSBS09} uses square waves to gate a SPAD and
a differencing circuit to subtract the output signals during two
consecutive clocks to acquire the weak avalanche signal. Each of the two
methods has its own advantages and disadvantages. SG has a simple
frequency spectrum and thus can be filtered. However, it is
significantly challenging to reduce the sine frequency response
to the pure electronic noise level using only filters.
Moreover, when the amplitude of background signal is highly
attenuated other frequency components, like the harmonics of the
fundamental frequency, can encumber the
minimization of the background signal. On the other hand, the rejection
ratio of the SD circuit is independent of frequency, which facilitates
the discrimination of weak avalanches. However, designing a
high-bandwidth and high-rejection differencing circuit is also quite challenging. As such,
both techniques require complicated and sophisticated electronics,
which can prevent their implementation for practical applications.

\begin{figure}[]
\centering
(a)
\includegraphics[scale=0.27]{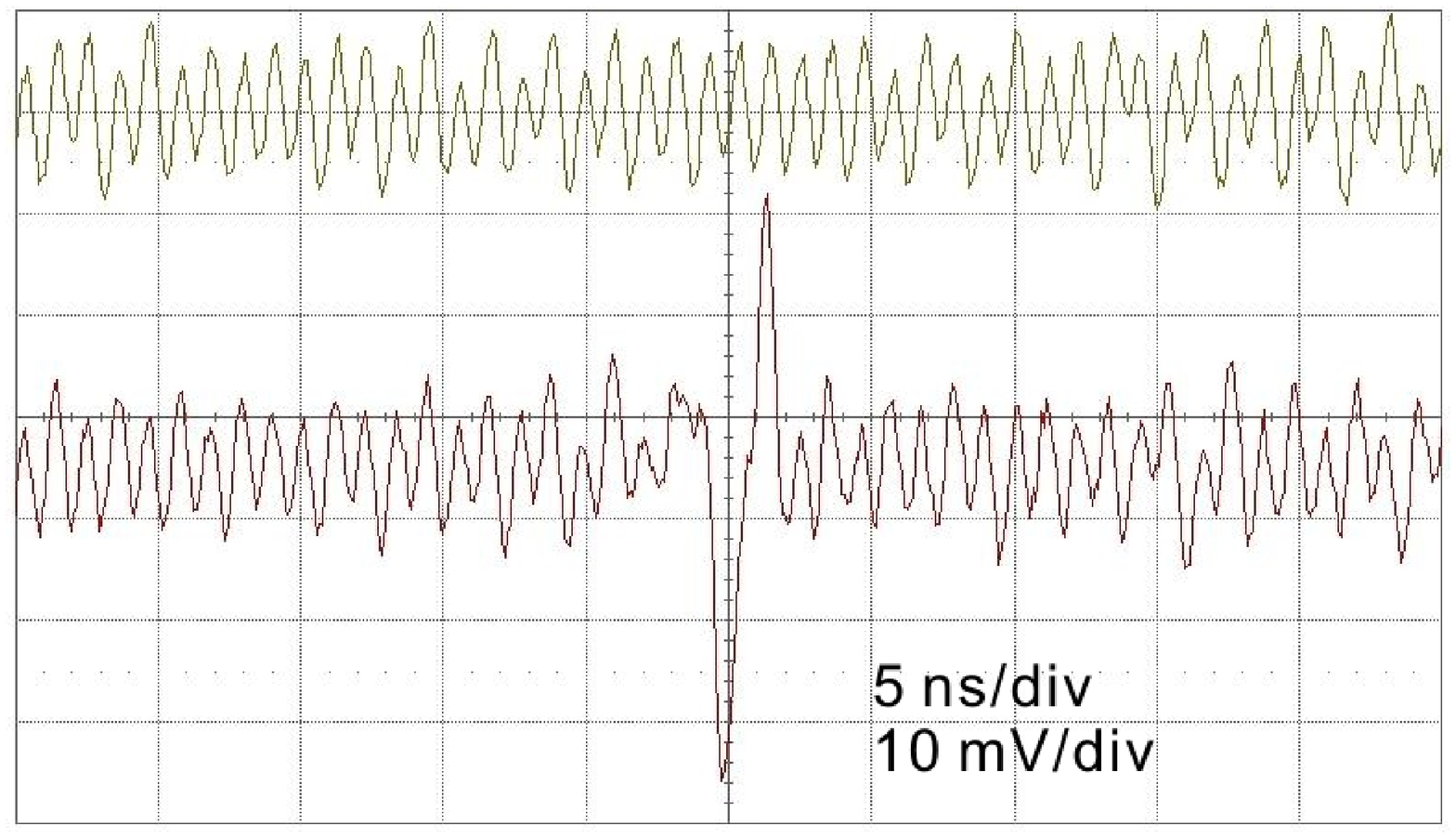}\\
(b)
\includegraphics[scale=0.27]{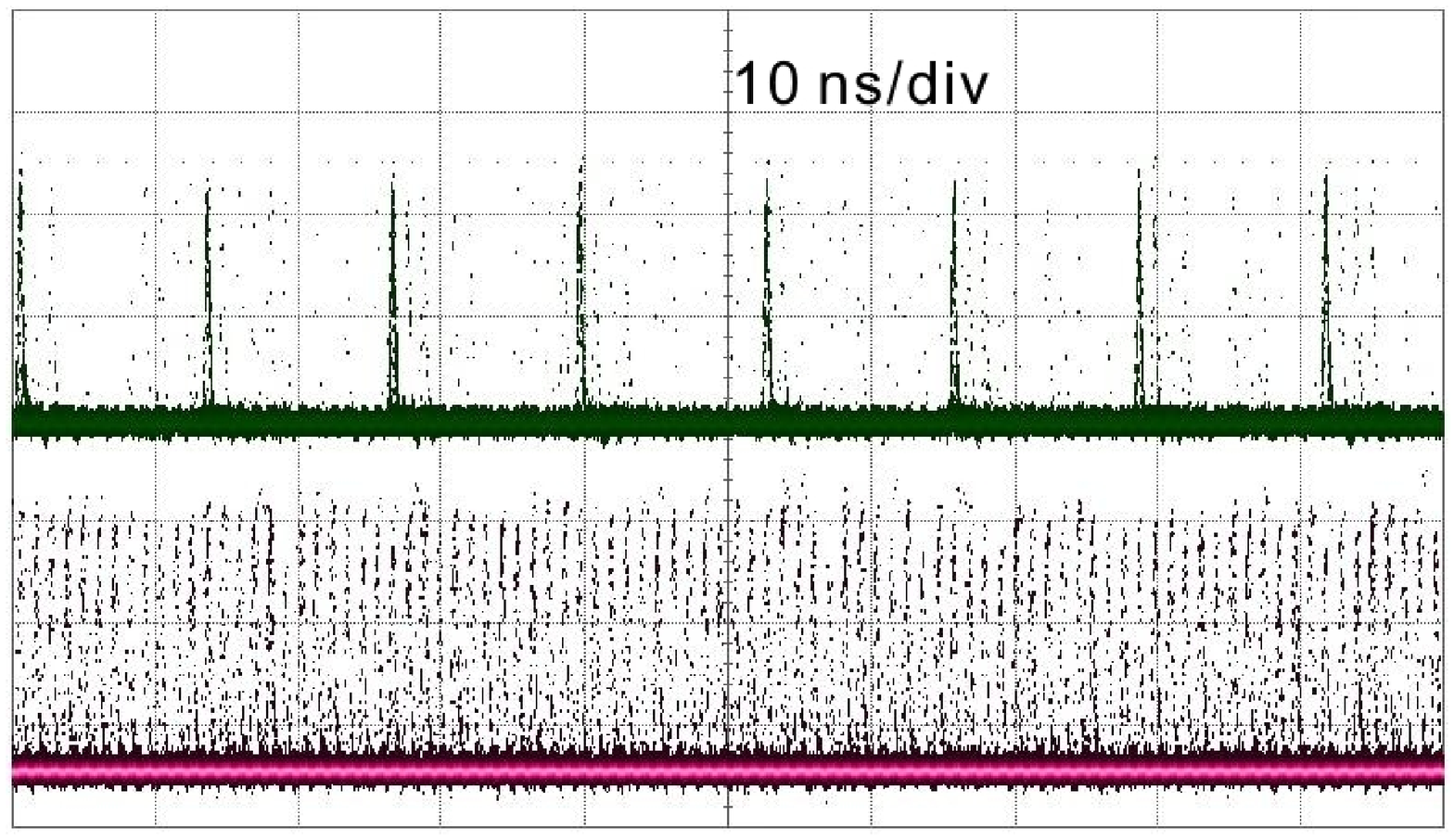}
\caption{(a) Typical output signals after a 20-dB high-bandwidth amplifier (amp2), showing the
amplitudes of background signal (the upper curve) and avalanche signal (the lower curve).
(b) The typically observed persistence of avalanche signals after a discriminator (dis).
The lower curve shows the random noise avalanches without
photon illumination, which are equally distributed in each gate. The upper curve shows the avalanches
with pulsed photon illumination, where most of the avalanches are created by photon absorption.
$f_{g}$ is 921\,MHz in both (a) and (b) while $f_{p}$ in (b) is $\sim$ 77\,MHz ($f_{g}$/12).}
\label{ava}
\end{figure}

\begin{figure}[]
\centering
\includegraphics[scale=0.37]{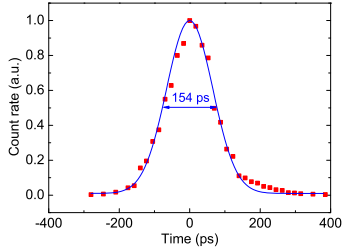}
\caption{Plot of the count rate as a function of the relative position of the laser pulse to the peak of gate.
The effective gating width (FWHM) is $\sim$ 154\,ps with an excess bias of 2\,V and $f_{g}$=921\,MHz.}
\label{delay}
\end{figure}

In this Letter, we report a simple and practical method for rapid gating
SPADs that combines aspects of the SG and SD approaches. The
overall implementation is easier than each of these previous
techniques independently as the requirements for each technique are
relatively unsophisticated.

The experimental setup is shown in Fig. \ref{sine}. The sine waves
from the synthesized signal generator (MG3601A, Anritsu) are
split by a 6-dB power divider. One part drives
the laser diode (LD: PicoQuant PDL 800-B, 30\,ps FWHM, max. 80\,MHz
repetition frequency) for the optical characterization.
The other part is amplified by a amplifier
(amp1, ZHL-42W, Mini-Circuits) and then coupled to the
anode of the SPAD via a 1\,nF capacitor. The gate signal typically
consists of a DC voltage ($V_{dc}$) of $\sim$ 55\,V and a $V_{pp}$
of $\sim$ 12\,V. The output from the cathode of
the SPAD is filtered by home-made BSFs,
rejecting the fundamental frequency $f_{g}$ and harmonics, especially
2$f_{g}$ and 4$f_{g}$, which are produced mainly due to the
nonlinear frequency response of the SPAD. After the BSFs, the background
amplitude is normally less than 40\,mV depending on the BSF adjustment.
The BSFs can contribute over 30\,dB of attenuation with the
remaining attenuation due to the voltage distribution by resistance.
Furthermore, the background signal can be suppressed down to the
electronic noise level by a self-differencing circuit.
A power divider first splits each pulse into two pulses.
The inverted pulse is then recombined with the preceding pulse
delayed by one clock. Finally, the background level
is less than 1\,mV while the avalanche level is around 2\,mV, see
Fig. \ref{ava}(a). The difference between the cases without and with
photon illumination (mean photon number per pulse $\mu$ $\sim$ 1,
laser frequency $f_{p}$=$f_{g}$/12), shown in Fig. \ref{ava}(b),
clearly illustrates the single-photon counting capability of our scheme.

\begin{figure}[]
\centering
\includegraphics[scale=0.48]{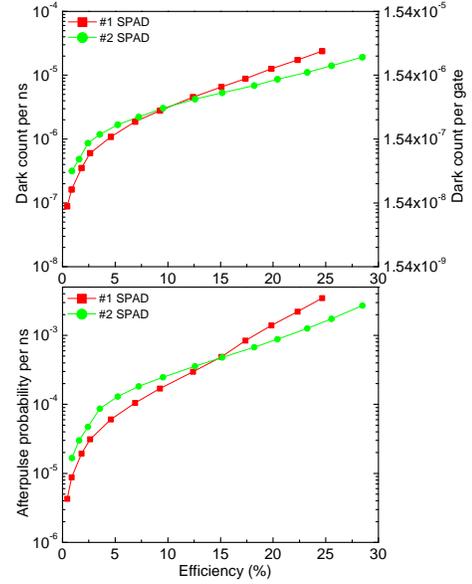}\\
\caption{Dark count (a) and afterpulse probabilities (b) vs detection efficiency for two SPADs, with
$f_{g}$=921\,MHz, $f_{p}=f_{g}/12$, $\mu$=0.1 and 10\,ns ``count-off time''.}
\label{parameter}
\end{figure}

We measure the parameters of two different SPADs, \#1 SPAD (JDSU0131E6739) and \#2 SPAD (JDSU0131E6738)
cooled to -30\,$^{\circ}$C. A 10\,ns ``count-off time'' is also applied,
which means that once an avalanche is triggered the avalanches during the following 10\,ns
won't be counted. This allows us to reduce the afterpulse probability and false electronic counts following an
avalanche. The effective gating width ($\Delta t$), shown in Fig \ref{delay}, is 154\,ps,
corresponding to a duty cycle of 14.2\,\%.

In Fig. \ref{parameter}, the efficiency ($\eta$) is calculated by,
\begin{equation}
\eta=1/\mu\times \ln((1-R_{dc}/f_{g})/(1-R_{de}^{c}/f_{p})),
\label{eff}
\end{equation}
considering a Poisson photon number distribution.
$R_{dc}$ is the dark count rate and $R_{de}^{c}$ is the coincidence rate between
detection and laser pulses. The dark count per ns ($P_{dc}^{ns}$)
is calculated by $P_{dc}^{ns}=R_{dc}/(f_{g}\Delta t)$, neglecting the afterpulsing of dark counts.
As shown in Table. \ref{list}, at $f_{g}$=921\,MHz and $\eta$=9.3\,\%,
$P_{dc}^{ns}$ is 2.8$\times10^{-6}$\,ns$^{-1}$ for \#1 SPAD, or 4.3$\times10^{-7}$ per gate,
which is very close to the parameter, 2.5$\times10^{-6}$\,ns$^{-1}$, measured at 10\,\% efficiency and -30\,$^\circ$C
in the conventional gating with the same SPAD.

\begin{table}[tp]
\caption{Parameter comparison between this method and the two other
techniques for rapid gating, as well as the active quenching
gated-mode (AQ).}
\begin{center}
\begin{tabular}{lllll}
  \hline
  \hline
   Parameter & This Letter & SD\cite{YKSS07} & SG\cite{NAI09} & AQ\cite{ZTGGZ09} \\
   \hline
   Temperature ($^\circ$C) & -30  & -30 & -50 & -35 \\
   $f_{g}$  & 921\,MHz  & 1.25\,GHz & 1.5\,GHz & 10\,kHz \\
   \boldmath{$\eta$} \textbf{(\%)} & 9.3  & 10.9 & 10.8 & 10.7 \\
   \boldmath{$P_{dc}^{ns}$\textbf{(}$\times 10^{-5}$\,\textbf{ns}$^{-1}$\textbf{)}} & 0.28 & 1.5 & 0.63 & 0.57 \\
   \hspace{10mm}{$\Delta t$}& 154\,ps  & 170\,ps & 100\,ps & 100\,ns \\
   \hspace{10mm}{$P_{ap}$ (\%)} & 3.4 & 6.2 & 2.8 & 1.8 \\
   \hspace{10mm}{$R_{de}$ (kHz)} & 732  & 213 & 108 & 1 \\
   \boldmath{$P_{ap}^{ns}$\textbf{(}$\times 10^{-5}$\,\textbf{ns}$^{-1}$\textbf{)}} & 16 & 6.3 & 2 & 18.3 \\
   \textbf{Deadtime} & 10\,ns & 10\,ns & 50\,ns & 15\,$\mu$s \\
  \hline
  \hline
\end{tabular}
\label{list}
\end{center}
\end{table}

From the relationship between $R_{de}^{c}$ and detection rate ($R_{de}$), the afterpulse
probability ($P_{ap}$) can be deduced as
\begin{equation}
P_{ap}=(R_{de}-R_{de}^{c}-11/12\times R_{dc})/R_{de}^{c}.
\label{ap}
\end{equation}
This implies that $P_{ap}$ highly depends on $R_{de}$.
In order to quantify and compare $P_{ap}$ under different conditions, we depict
the normalized parameters to ns$^{-1}$ as shown in Fig. \ref{parameter}.
The best and direct solution for evaluating the afterpulse probability per ns ($P_{ap}^{ns}$)
is to use the double-gate method \cite{ZTGGZ09}. However, it is quite difficult to directly apply such
a method for rapid gating systems. An alternative solution is to divide $P_{ap}$ by the
average effective time between detections,
\begin{equation}
P_{ap}^{ns}=P_{ap}/(f_{g}\Delta t/R_{de})\sim P_{ap}f_{p}\mu\eta/(f_{g}\Delta t),
\label{norm}
\end{equation}
where the interval time is $\sim\mu$s level for 10\,\% efficiency, see Table. \ref{list}, and
therefore the deadtime is negligible. If the interval time is much less than the detrapping
lifetime of afterpulses \cite{ZTGGZ09}, $P_{ap}^{ns}$ can well describe the afterpulsing behaviors,
otherwise long interval times or small detection rates will underestimate $P_{ap}^{ns}$.
For comparison, we also take the data from Ref. \cite{YKSS07} and Ref. \cite{NAI09}
and calculate them according to Eq. \ref{norm}, and list the results in Table. \ref{list}.
In our case, we measure a $P_{ap}$ of 3.4\,\% and calculate a $P_{ap}^{ns}$ of 1.6$\times10^{-4}$\,ns$^{-1}$, which
is larger than those in SD and SG, since our value of $R_{de}$
is much higher than theirs, but still comparable to the value using the active quenching
with 15\,$\mu$s deadtime \cite{ZTGGZ09}. In general, $P_{ap}^{ns}$ is larger than $P_{dc}^{ns}$,
which implies that the afterpulsing still dominates the noise
characteristics of SPADs in the rapid gating.

We also characterize the count rate behavior. As $\mu$ rises count
rate increases linearly when $\mu<$10 and finally the count rate is
saturated close to $f_{p}$. The theoretically maximum count rate can
approach 100\,MHz due to 10\,ns ``count-off time''.

Rapid gating is highly suited for applications requiring high-speed synchronized single-photon detection
like short-distance and high rate QKD \cite{NFIHT07,QKD1,QKD2}, but for the applications of asynchronous or
low photon flux detection, such as the long-distance QKD,
the free-running detector, which has low noise characteristic with large deadtime,
appears to be a better choice \cite{TSGZR07,ZTGGZ09}.

In summary, we have implemented a simple and practical method for
high-speed near infrared single-photon detection based on InGaAs/InP
SPADs.

The authors would like to thank J.-D. Gautier for technical assistances,
as well N. Gisin, O. Guinnard, A. Rochas, Z.L. Yuan and N. Namekata for
useful discussions. We also acknowledge financial support from the
Swiss NCCR - Quantum Photonics.

\end{document}